\documentclass{ws-procs975x65}

\def\zone{the error due to the even zonal harmonics of geopotential\ }

\def\rfr#1{eq.(\ref{#1})}

\def\leti{Lense--Thirring}

\def\eqi{\begin{equation}}
\def\eqf{\end{equation}}
\def\eqia{\begin{eqnarray}}
\def\eqfa{\end{eqnarray}}

\def\rp#1#2{{#1\over#2}}

\def\mlt{{\rm \mu_{LT}}}

\def\lb#1{\label{#1}}
\def\lg{LAGEOS}
\def\lgg{LAGEOS II}

\begin{document}

\def\nocropmarks{\vskip5pt\phantom{cropmarks}}

\let\trimmarks\nocropmarks

\markboth{Lorenzo Iorio} {The new Earth gravity models and the
measurement of the Lense--Thirring effect}

\catchline{}{}{}

\title{THE NEW EARTH GRAVITY MODELS AND THE MEASUREMENT OF THE LENSE--THIRRING EFFECT}

\author{\footnotesize LORENZO IORIO}

\address{Dipartimento di Fisica dell'Universit$\grave{ a}$ di Bari\\
Via Amendola 173, 70126, Bari, Italy}

\maketitle

\abstracts{We examine how the new forthcoming Earth gravity models
from the CHAMP and, especially, GRACE missions could improve the
measurement of the general relativistic Lense--Thirring effect
according to the various kinds of observables which could be
adopted. In a very preliminary way, we use the recently released
EIGEN2 CHAMP--only and GRACE01S GRACE--only Earth gravity models
in order to assess the impact of the mismodelling in the even
zonal harmonic coefficients of geopotential which represents one
of the major sources of systematic errors in this kind of
measurement.}

\section{Introduction}
The post--Newtonian gravitomagnetic Lense--Thirring
effect$^{1,2,3}$, in the slow motion and weak field approximation
of the General Theory of Relativity, consists of secular
precessions of the longitude of the ascending node $\Omega$ and
the argument of pericentre\footnote{In the original paper by Lense
and Thirring the longitude of the pericentre
$\varpi=\Omega+\omega$ is used instead of $\omega$. } $\omega$ of
the orbit of a test particle freely falling in the gravitational
field of a central spinning object of mass $M$ and proper angular
momentum $J$
\begin{eqnarray}
\dot\Omega_{\rm LT} &=&\rp{2GJ}{c^2 a^3(1-e^2)^{\rp{3}{2}}},\\
\dot\omega_{\rm LT} &=&-\rp{6GJ\cos i}{c^2
a^3(1-e^2)^{\rp{3}{2}}},
\end{eqnarray} where $a,\ e$ and $i$ are the semimajor axis, the eccentricity and the inclination, respectively, of
the orbit of the test particle and $G$ is the Newtonian
gravitational constant; $c$ is the speed of light in vacuum.
\subsection{The LAGEOS--LAGEOS II Lense--Thirring experiment}
The first attempts to measure such effect in the gravitational
field of Earth are due to Ciufolini and coworkers which analysed
the data of the orbits of the existing LAGEOS and LAGEOS II
geodetic satellites collected with the Satellite Laser Ranging
(SLR) technique$^{4}$. The observable used is the following
combination$^{5}$ of the orbital residuals of the nodes of LAGEOS
and LAGEOS II and the perigee of LAGEOS II
\eqi\delta\dot\Omega^{\rm LAGEOS }+c_1\delta\dot\Omega^{\rm
LAGEOS\ II}+c_2\delta\dot\omega^{\rm LAGEOS\ II}\sim 60.2\mu_{\rm
LT},\lb{ciufcomb}\eqf with \eqi c_1= 0.304,\ c_1=
-0.350,\lb{ciufcoeef}\eqf
\begin{table}[htbp]
\ttbl{30pc}{Orbital parameters of LAGEOS, LAGEOS II, LARES, S1 and
S2.}
{\begin{tabular}{lccccc}\\
\multicolumn{6}{c}{}\\[6pt]\hline
Orbital parameter & \lg & \lgg & LARES & S1 & S2\\ \hline
$a$ (km) & 12270 & 12163 & 12270 & 12000 & 12000\\
$e$ & 0.0045 & 0.014 & 0.04 & 0.05 & 0.05\\
$i$ (deg) & 110 & 52.65 & 70 & 63.4 & 116.6\\
\hline
\end{tabular}}
\end{table}\label{uffaa}
The quantity $\mu_{\rm LT}$ is a solved--for least square
parameter which is 0 in Newtonian mechanics and 1 in the General
Theory of Relativity. The gravitomagnetic signature is a linear
trend with a slope of 60.2 milliarcseconds per year (mas yr$^{-1}$
in the following). The claimed total accuracy is of the order of
20$\%$--30$\%$. In Table 1 we quote the orbital parameters of the
existing or proposed LAGEOS--like satellites useful for general
relativistic tests.

The main sources of systematic errors in this experiment are
\begin{itemize}
\item
the unavoidable aliasing effect due to the mismodelling in the
classical secular precessions induced on $\Omega$ and $\omega$ by
the even ($\ell =2,\ 4,\ 6,...$) zonal ($m=0$)
coefficients\footnote{$J_{\ell}\equiv -C_{\ell 0}$, $\ell = 2,\
4,\ 6,...$. The unnormalized Stokes coefficients $C_{\ell m}$ of
degree $\ell$ and order $m$ can be obtained from the normalized
Stokes coefficients $\overline{C}_{\ell m}$ according to $C_{\ell
m}= N_{\ell m}\overline{C}_{\ell m}$ where $N_{\ell m
}=\left[\rp{(2\ell +1)(2-\delta_{0m})(\ell - m)!}{(\ell +
m)!}\right]^{\rp{1}{2}}$} $J_{\ell}$ of the multipolar
expansion\footnote{It accounts for the oblateness of Earth
generated by its diurnal rotation.} of the terrestrial
gravitational field
\item
the non--gravitational perturbations affecting especially the
perigee of LAGEOS II. Their impact on the proposed measurement is
difficult to be reliably assessed
\end{itemize}
It turns out that the mismodelled classical precessions due to the
first two even zonal harmonics of the geopotential $J_2$ and $J_4$
are the most insidious source of error in the measurement of the
Lense--Thirring effect with the nodes of LAGEOS and LAGEOS II and
the perigee of LAGEOS II only. It is so also because of their
secular variations $\dot J_2$, $\dot J_4$, which can be accounted
for by an effective $J_2$ coefficient $\dot J_2^{\rm eff}\sim\dot
J_2+0.371\dot J_4+0.079\dot J_6+0.006\dot J_8-0.003\dot
J_{10}...$, and of the fact that the Lense-Thirring effect is
embedded in their recovered values released by any Earth gravity models
since the gravitomagnetic acceleration is a priori included in the
force models used in obtaining, among other things, the even zonal
coefficients. The combination of \rfr{ciufcomb} is insensitive
just to $J_2$ and $J_4$. According to the full covariance matrix
of the EGM96 gravity model$^6$, the error due to the remaining
uncancelled even zonal harmonics amounts to almost$^7$ 13$\%$.
However, if the correlations among the even zonal harmonic
coefficients are neglected and the variance matrix, in a
root--sum--square fashion, is used\footnote{Such approach is
considered more realistic by some authors$^8$ because nothing
would assure that the correlations among the even zonal harmonics of
the covariance matrix of the EGM96 model, which has been obtained
during a multidecadal time span, would be the same during an
arbitrary past or future time span of a few years as that used in
the LAGEOS--LAGEOS II
\leti\ experiment or in the proposed LAGEOS--LARES mission.
Moreover, in the same paper it is claimed that a correct and
reliable evaluation of the impact of the non-gravitational
perturbations of the perigee of LAGEOS II on the proposed
measurement is rather troublesome.}, \zone amounts to$^7$ 45$\%$.
With this estimate the total error of the LAGEOS--LAGEOS II
\leti\ experiment would be of the order of 50$\%$. If the sum of
the absolute values of the individual errors is considered, an
upper bound of 83$\%$ on the systematic error due to the even
zonal harmonics is obtained; the total error in the LAGEOS--LAGEOS
II
\leti\ experiment would, then,
amount to almost 100$\%$. It must be considered that, since in
EGM96 a strong correlation among the various recovered even zonal
coefficients  does occur, the approach of the sum of the absolute
values of the individual errors would, probably, yield a more realistic
estimate of this kind of systematic error.
\subsection{The supplementary orbital planes option}
The originally proposed LARES mission$^9$  consists of the launch
of a LAGEOS--type satellite--the LARES--with the same orbit of
LAGEOS except for the inclination $i$ of its orbit, which should
be supplementary to that of LAGEOS, and the eccentricity $e$,
which should be one order of magnitude larger. The choice of the
particular value of the inclination for LARES is motivated by
the fact that in this way, by using as observable the sum of the
nodes of LAGEOS and LARES
\eqi\Sigma\delta\dot\Omega\equiv\dot\Omega^{\rm LAGEOS}+\delta\dot\Omega^{\rm
LARES},\lb{nodisomma}\eqf it should be possible to cancel out, to
a high degree of accuracy, all the contributions of the even zonal
harmonics of the geopotential, which depends on $\cos i$, and add
up the Lense--Thirring precessions which, instead, are independent
of $i$. The systematic error due to the even zonal coefficients of
geopotential would range from$^{10}$ 0.8$\%$ to 4.6$\%$ for an
uncertainty in $i_{\rm LARES}$ up to 1 deg with respect to its
nominal values of Table 1, according to the variance matrix of
EGM96 used in a root--sum--square fashion. If, instead, the sum of
the absolute values of the individual errors is considered, the
systematic error due to the even zonal coefficients of
geopotential would amount to almost 2$\%$ for $i_{\rm LARES}=70$
deg and to $10\%$ for a departure of $i_{\rm LARES}$ of 1 deg from
its nominal value. The impact of the non-gravitational perturbations would be
far less important because the nodes of the LAGEOS-like satellites
are almost insensitive to them.

Since the Lense--Thirring precession of the pericentre depends on
$\cos i$ and the classical precessions due to the even zonal
harmonics of geopotential depend on $\cos^2 i$ and odd powers of
$\sin i$, also the difference of the perigees$^{11, 12}$
\eqi\Delta\dot\omega\equiv\delta\dot\omega^{\rm S1}-\delta\dot\omega^{\rm
S2},\lb{peridiff}\eqf of a pair of satellites with identical
orbits in supplementary orbital planes could be used, in priciple, in order to
measure the Lense--Thirring effect. It has been shown that, for a
pair of LAGEOS--like satellites S1 and S2 (see Table 1) with the
same physical properties of the LAGEOS satellites, the total error
would not be less than$^{12}$ 5$\%$, mainly due to the
non--gravitational perturbations.
\subsection{Other approaches}
In order to cope with the problem of the impact of the unavoidable
orbital injection errors in the inclination of LARES on the
systematic error due to the even zonal harmonics of geopotential,
the following observable has also been proposed$^{13}$
\eqi \delta\dot\Omega^{\rm LAGEOS}+c_1\delta\dot\Omega^{\rm
LAGEOS\ II}+c_2\delta\dot\Omega^{\rm
LARES}+c_3\delta\dot\omega^{\rm LARES} \ \sim
61.8\mlt,\lb{combinopg}\eqf with \eqi c_1  \sim 3\times 10^{-3},\
c_2  \sim  9.9\times 10^{-1},\ c_3  \sim  1\times
10^{-3}\lb{optiscomb}. \eqf
It would allow to neglect the problem of the correlations between
the various even zonal coefficients of geopotential. Moreover, the
secular variations in the even zonal harmonic coefficients of
geopotential would not affect such observable. The same would also
hold  for the bias of the Lense--Thirring effect itself on the
even zonal harmonics because it is mainly concentrated in $J_2$
and $J_4$. Notice that such an observable does not include the
perigee of LAGEOS II.

Another possible way to overcome the uncertainties related to the
error budget evaluation due to the non--gravitational
perturbations which affects the perigee of LAGEOS II by using the
existing satellites is the following combination$^{14}$
\eqi\delta\dot\Omega^{\rm LAGEOS }+c_1\delta\dot\Omega^{\rm
LAGEOS\ II}\sim 48.2\mu_{\rm LT},\ c_1=0.546\lb{nodi1e2}\eqf
A similar approach is implicitly presented in ref.$^{15}$. The
observable of \rfr{nodi1e2} would be affected neither by the
secular variations in the $J_{\ell}$ coefficients nor, at least to
a certain extent, by the a priori Lense--Thirring bias of the even
zonal harmonics.
\subsection{The role of the various Earth gravity models}
Here we wish to discuss the impact of the new, forthcoming Earth
gravity models from the CHAMP$^{16}$ and GRACE$^{15}$ missions,
which should greatly improve our knowledge of the terrestrial
gravitational field, on the different combinations used or proposed for the
measurement of the Lense--Thirring effect. Very recently, the
EIGEN2$^{17}$, GRACE01S$^{18}$ and GGM01C/S\footnote{It can be
retrieved on the WEB at http://www.csr.utexas.edu/grace/gravity/.
The GRACE--only GGM01S model was combined with the TEG-4
information equations (created from historical multi--satellite
tracking data; surface gravity data and altimetric sea surface
heights) to produce the preliminary gravity model GGM01C. } models
from the CHAMP and GRACE only data, respectively, have been
released. Although very preliminary and lacking of reliable
validation and calibration tests, they can be used in order to get
an idea of which the improvements in the Lense--Thirring
tests could be. In Table 2 we quote the sigmas of the normalized even
zonal Stokes coefficients $\sigma_{\overline{C}_{\ell 0}}$ for
EGM96, EIGEN2 and GRACE01S. Note that for the GRACE models the
sigmas are not the formal errors but are approximately calibrated.
For EIGEN2 the formal errors are released: it is likely that they
are rather optimistic, at least for the low degree even zonal
harmonics up to $\ell=20^{17}$ .
\begin{table}[htbp]
\ttbl{30pc}{Errors in the even zonal normalized Stokes
coefficients $\sigma_{\overline{C}_{\ell 0}}$ for various Earth
gravity models up to degree $\ell=20$. The GGM01C model is very
similar to GRACE01S, apart from
$\sigma_{{\bar{C}}_{20}}=1.1677\times 10^{-11}$ and
$\sigma_{{\bar{C}}_{40}}=1.7874\times 10^{-11}$.}
{\begin{tabular}{lccc}\\
\multicolumn{4}{c}{}\\[6pt]\hline
$\ell$ & EGM96 & EIGEN2 & GRACE01S\\
\hline
2 & 3.561$\times 10^{-11}$ & 1.939$\times 10^{-11}$ & 3.106$\times 10^{-10}$\\
4 & 1.042$\times 10^{-10}$ & 2.230$\times 10^{-11}$ & 2.713$\times 10^{-11}$\\
6 & 1.449$\times 10^{-10}$ & 3.136$\times 10^{-11}$ & 9.568$\times 10^{-12}$\\
8 & 2.266$\times 10^{-10}$ & 4.266$\times 10^{-11}$ & 6.428$\times 10^{-12}$\\
10 & 3.089$\times 10^{-10}$ & 5.679$\times 10^{-11}$ & 7.180$\times 10^{-12}$\\
12 & 4.358$\times 10^{-10}$ & 7.421$\times 10^{-11}$ & 5.022$\times 10^{-12}$\\
14 & 5.459$\times 10^{-10}$ & 9.561$\times 10^{-11}$ & 4.376$\times 10^{-12}$\\
16 & 5.313$\times 10^{-10}$ & 1.216$\times 10^{-10}$ & 3.805$\times 10^{-12}$\\
18 & 4.678$\times 10^{-10}$ & 1.532$\times 10^{-10}$ & 4.412$\times 10^{-12}$\\
20 & 4.690$\times 10^{-10}$ & 1.914$\times 10^{-10}$ & 3.571$\times 10^{-12}$\\
\hline
\end{tabular}}
\end{table}
\label{cazzo} It should be noted that $J_2$ is known much less
accurately in GRACE01S than in EGM96 and EIGEN2.
\section{Combinations with the existing or proposed geodetic satellites of LAGEOS--type}
The classical secular precessions of the node and the perigee, for
a given even zonal harmonic of degree $\ell$, are proportional to
$R^{\ell}a^{-\left(\rp{3+2\ell}{2}\right)}$, where $R$ is the
Earth mean equatorial radius. Then, the relevant classical secular
precessions of the node and the perigee of the LAGEOS satellites,
due to their high altitude (see Table 1), are those induced just
by the first few five--six even zonal harmonics of geopotential in
the sense that the error in the Lense--Thirring measurement due to
geopotential does not change if the even zonal harmonics of degree
higher than $\ell=10$--12 are neglected in the calculations.
\subsection{The full--range even zonal harmonics observables}
Let us recall that the sum of the nodes $\Sigma\dot\Omega$ of
LAGEOS and LARES (and of S1 and S2), and the difference of the
perigees $\Delta\dot\omega$ of S1 and S2 do not cancel any even
zonal harmonic coefficients of geopotential due to the unavoidable
orbital injection errors; in the case of the originally proposed
LAGEOS-LARES mission this non perfect cancellation of the effects
of the even zonal harmonics would occur even if the orbital
parameters of LARES were exactly the same as in Table 1 because
$e_{\rm LARES}=0.04$, while $e_{\rm LAGEOS}=0.0045$. This implies
that it is of the utmost importance that the new gravity models
yield notable improvements especially for $J_2$, $J_4$ and, to a
lower extent, $J_6$ and $J_8$ which would affect such observables
and which are the most relevant sources of systematic errors for
the Lense--Thirring precessions of the nodes and the perigees of
each satellite. This is also the case for the combination of
\rfr{nodi1e2} which involves the node of LAGEOS and the node of
LAGEOS II; it cancels out the first even zonal harmonic $J_2$ and
is affected by the remaining $J_4$, $J_6$,...

Let us consider the sum of the nodes of LAGEOS and LARES. By
assuming an orbital injection error in $i_{\rm LARES}$ up to 1
deg, it turns out that the variance matrix of EIGEN2 up to degree
$\ell=20$ yields an error due to the even zonal harmonics of
geopotential spanning from 0.2$\%$ to 1.6$\%$. GRACE01S, instead,
induces an error which ranges from 0.2$\%$ to 24$\%$. The range
error induced by GGM01C is 0.2$\%$-0.9$\%$. Let us recall that the
error due to EGM96 ranges from 2$\%$ to\footnote{Note that for
EIGEN2, GRACE01S and GGM01C a root--sum--square calculation with
their variance matrices yields realistic estimates because,
contrary to EGM96, their covariance matrices are almost diagonal.}
$10\%$. These numbers can be explained if we consider that for
large departures from the condition of ideal supplementary orbital
planes the contributions of all the even zonal harmonics--and,
especially, those of lower degrees--affect the systematic error
due to geopotential. From Table 2 it can be noticed that $J_2$ is
far better known in EIGEN2 than in GRACE01S. An analysis of the
contributions of the various even zonal harmonics in the building
up of the systematic error due to geopotential shows that for
departures of $i_{\rm LARES}$ of 1 deg from its nominal value of
Table 1 the most important role is played just by $J_2$. This
also explains the better results obtainable with GGM01C.

The combination of \rfr{nodi1e2} is affected by the uncancelled
even zonal harmonics $J_4$, $J_6$,... at a 177$\%$, 22$\%$, 21$\%$
and $14\%$ level according to the sum of the absolute values of
the errors of EGM96 and to the variance matrices of EIGEN2,
GRACE01S and GGM01C, respectively. The small difference between
the EIGEN2 and GRACE01S results can be explained by noticing that
$J_2$ does not affect \rfr{nodi1e2}, the error in $J_4$ is
slightly better in EIGEN2 than in GRACE01S and the accuracy of the
GRACE01S determinations of $J_6$, $J_8$,... is only slightly
better than that of EIGEN2. The better results for GGM01C,
especially for $J_4$, explain the result obtained with such Earth gravity
model.
\subsection{The partial--range even zonal harmonics observables}
If, instead, we consider the suitably designed combinations of $N$
orbital residuals of the LAGEOS satellites which cancel out the
first $N-1$ even zonal harmonics\footnote{We assume $N>2$.} the
situation is different in the sense that the improvements of the
remaining higher degree even zonal harmonics become relevant in
order to reduce the systematic error due to the even zonal
harmonics of geopotential which, for $N>3$ is well below 1$\%$
also with EGM96 (RSS calculation).

Let us consider the currently used observable of \rfr{ciufcomb}
which cancels out $J_2$ and $J_4$. According to the variance
matrices of EGM96, EIGEN2 and GRACE01S up to degree $\ell=20$ the
systematic error due to geopotential amounts to 83$\%$ (sum of the
absolute values of the errors), 9$\%$ and 2$\%$ (root--sum--square
calculations), respectively. This shows the importance of reducing
the uncertainties in $J_6$ and, to a lower extent, $J_8$.

The combination of \rfr{combinopg} is affected by the remaining
even zonal harmonics $J_8$, $J_{10},$... far less than 1$\%$.
Indeed, for a maximum orbital injection error in $i_{\rm LARES}$
of 1 deg the systematic error due to geopotential amounts to
2.2$\%$--3$\%$ according to the sum of the absolute values of the
errors in the variance matrix of EGM96 up to degree $\ell=20$, to
0.08$\%$--0.1$\%$ according to the variance matrix of EIGEN2, used
in a root--sum--square fashion, up to degree $\ell=20$ and to
0.007$\%$--0.01$\%$ according to the variance matrix of GRACE01S,
used in a root--sum--square fashion, up to degree $\ell=20$. This
can be explained by noticing from Table 2 that the higher degree
even zonal harmonics are far better determined in GRACE01S (and
GGM01C/S) than in the other Earth gravity models.
\section{Combinations with the other existing geodetic satellites}
In principle, a very appealing possibility would be the use of
combinations which include also the orbital elements of the other
existing geodetic satellites like Ajisai, Starlette, Stella,
WESTPAC1, ETALON1, ETALON2, apart from those of LAGEOS and LAGEOS
II; this option has been studied in some previous works$^{19}$. In
Table \ref{altrisat} the orbital elements of the other existing
SLR satellites are reported.
\begin{table}[htbp]
\ttbl{30pc}{Orbital parameters of Ajisai, Starlette, Stella,
WESTPAC1, ETALON1 and ETALON2}
{\begin{tabular}{lcccccc}\\
\multicolumn{7}{c}{}\\[6pt]\hline
Orbital parameter & Ajisai & Starlette & Stella & WESTPAC1 &
ETALON 1 & ETALON2\\ \hline
$a$ (km) & 7870 & 7331 & 7193 & 7213 & 25498 & 25498\\
$e$ & 0.001 & 0.0204 & 0 & 0 & 0.00061 & 0.00066\\
$i$ (deg) & 50 & 49.8 & 98.6 & 98 & 64.9 & 65.5\\
\hline
\end{tabular}}
\end{table}\label{altrisat}
For example, one might think about some combinations including
only the nodes of the LAGEOS satellites and of some of the other
geodetic SLR satellites. Unfortunately, it turns out that this way
is unpracticable because the other satellites orbit at lower
altitudes than LAGEOS and LAGEOS II (see Table \ref{altrisat}), so
that they are sensitive to much more even zonal harmonics,
especially Starlette, Stella and WESTPAC1. Also EIGEN2 and
GRACE01S do not alter this situation in a significative way.
Indeed, it turns out that, while the systematic error due to the
even zonal harmonics of geopotential of a combination which
includes the nodes of LAGEOS and LAGEOS II, the perigee of LAGEOS
II and the node of Ajisai is sensitive only to the harmonics up to
degree $\ell=14$, according to the variance matrix of GRACE01S,
for another combination similar to the previous one,  with the
node of Starlette instead of the perigee of LAGEOS II, the error
induced by geopotential is sensitive well up to degrees
$\ell=40$--50. Moreover, the calculation of the classical secular
nodal precessions for degrees as high as $\ell=40$ is rather
unreliable because of instabilities of the numerical results.
However, it may be interesting to consider the following
combination \eqi\delta\dot\Omega^{\rm
LAGEOS}+c_1\delta\dot\Omega^{\rm LAGEOS\
II}+c_2\delta\dot\Omega^{\rm Ajisai}+c_3\delta\dot\Omega^{\rm
Starlette}+c_4\delta\dot\Omega^{\rm Stella}\sim 57.4\mu_{\rm
LT},\lb{multico}\eqf with $c_1=4.174,\ c_2=-2.705,\ c_3=1.508,\
c_4=-0.048$. According to a Root--Sum--Square calculation with the
variance matrix of GGM01C up to\footnote{It has been checked that
the error due to the even zonal harmonics remains stable if other
even zonal harmonics are added to the calculation. Moreover, it
turns also out that, up to $\ell=42$, there are no appreciable
fluctuations in the calculated classical secular precessions.}
$\ell=42$ the impact of the remaining even zonal harmonics of
degree $\ell\geq 10$ amounts to 21.6 mas yr$^{-1}$ which yields a
$37.6\%$ percent error due to geopotential in the measurement of
the Lense-Thirring effect with \rfr{multico}. The upper bound due
to the sum of the absolute values of the individual errors amounts
to 123$\%$. If the future GRACE--based gravity solutions will
improve the high degree ($J_{10},\ J_{12},\ J_{14},...$) even
zonal harmonics more than the low degree ($J_2,\ J_4,\ J_6,\ J_8$)
ones, the combination of \rfr{multico} might deserve some interest
in alternative to that of \rfr{nodi1e2}.

In conclusion, also with the new forthcoming Earth gravity models,
it is unlikely that the combinations involving the orbital
elements of the other existing geodetic satellites as well will
represent a real alternative to those which use only the existing
LAGEOS satellites and the proposed LARES. If, instead, we refer to the
existing LAGEOS satellites only, it may happen that the inclusion
of the nodes of the other geodetic satellites could become a
viable possibility.
\section{Conclusions}
The choice of the optimal observable for  measuring the
Lense--Thirring effect is dictated by the need of getting a
compromise among the reduction of the systematic gravitational
error due to geopotential, the systematic non--gravitational error
induced by the perturbations affecting especially the perigees of
the geodetic SLR satellites and the reliability of the SLR data of
the satellites orbits.

In regard to those requirements, it turns out that even with the
new, forthcoming Earth gravity models the multisatellite approach
involving the orbital elements of the existing SLR satellites
other than the LAGEOS ones should not be competitive with the
observables built up with the LAGEOS--type satellites only.

Among them, notice that the total accuracy obtainable with the
originally proposed combination by Ciufolini, which involves the
nodes of both LAGEOS and LAGEOS II and the perigee of LAGEOS II,
strongly depends on the correct evaluation of the impact of the
non--gravitational perturbations on the perigee of LAGEOS II
because it is arguable that the error due to geopotential could be
reduced to 1$\%$ or, perhaps, less in the near future. On the
contrary, for the combination proposed explicitly in ref.$^{14}$, which
involves the nodes of LAGEOS and LAGEOS II, the improvements in
the Earth gravity field even zonal coefficients, especially $J_4$,
$J_6$, $J_8$, will be of crucial importance in reducing the total
error to less than 10$\%$ because the impact of the
non--gravitational perturbations on the error budget is quite
negligible. Also a combination involving the nodes of LAGEOS,
LAGEOS II, Ajisai, Starlette and Stella might become interesting,
provided that notable improvements in the higher degree even zonal
harmonics $J_{10},\ J_{12},...$ will occur.

For the observables involving the proposed LARES, the
multisatellite combination including also the elements of LAGEOS
and LAGEOS II would not require any particular improvements in the
knowledge of the higher degree even zonal harmonics of
geopotential because the systematic error induced by them,
according to the present--day level of knowledge of the
terrestrial gravitational field, would be sufficiently small.

\end{document}